\documentclass[a4paper]{article}
\usepackage{Odyssey2024}
\usepackage{epsfig,amssymb,amsmath}
\usepackage{multirow, multicol}
\ninept
\usepackage{float}
\usepackage[caption = false]{subfig}
\usepackage{graphicx}
\title{Double Multi-Head Attention Multimodal System for Odyssey 2024 Speech Emotion Recognition Challenge}

\makeatletter
\def\name#1{\gdef\@name{#1\\}}
\makeatother
\name{{\em Federico Costa \textsuperscript{1}, Miquel India \textsuperscript{1}, Javier Hernando \textsuperscript{2,1}}}

\address{\textsuperscript{1}TALP Research Center, Universitat Politècnica de Catalunya (UPC), Spain. \\
\textsuperscript{2}Barcelona Supercomputing Center (BSC), Spain. \\
{\small \tt federico.costa@upc.edu, miquel.angel.india@upc.edu, javier.hernando@bsc.es} }
\begin{document}
\maketitle

\begin{abstract}
As computer-based applications are becoming more integrated into our daily lives, the importance of Speech Emotion Recognition (SER) has increased significantly. 
Promoting research with innovative approaches in SER, the Odyssey 2024 Speech Emotion Recognition Challenge was organized as part of the Odyssey 2024 Speaker and Language Recognition Workshop. 
In this paper we describe the Double Multi-Head Attention Multimodal System developed for this challenge.
Pre-trained self-supervised models were used to extract informative acoustic and text features. 
An early fusion strategy was adopted, where a Multi-Head Attention layer transforms these mixed features into complementary contextualized representations.
A second attention mechanism is then applied to pool these representations into an utterance-level vector.
Our proposed system achieved the third position in the categorical task ranking with a 34.41\% Macro-F1 score, where 31 teams participated in total.
\end{abstract}

\section{Introduction}

Natural Human-Computer Interaction (HCI) is becoming more and more important as machines and computer-based applications are gaining significance in our daily lives \cite{SER-2018}.
A key step in HCI is to recognize and address people's emotional states.
In addition, automatic emotion detection and classification are essential for automated tutoring systems, call centers, gaming, personal assistants and more. 
For these reasons, progress in these technologies can enable transformative applications in diverse areas such as healthcare, security and defense, education and entertainment.
Humans use various modalities to express their emotions, such as speech, text, facial expressions or gestures.
Because of the subtle expressive behaviors that arise during human interactions, automatic emotion recognition from speech in realistic domains is still a challenging task \cite{TEA-2013}.

Speech is essential for expressing a person's emotional state through prosody and/or paralinguistic context.
In the domain of Speech Emotion Recognition (SER), Machine Learning (ML) models have been constructed relying on the utilization of hand-crafted features, such as Mel-frequency Cepstral Coefficients (MFCC), pitch, energy, entropy or zero-crossing rate \cite{ESR-2006}.
The correlation between different emotions and specific speech features is still uncertain, and ongoing research is being conducted to investigate specific speech representations that can effectively capture human emotions' patterns.
However, deep learning-based models do not require the extraction of a large set of hand-crafted features because they can learn the features that are relevant to the task from spectrograms or even directly from raw waveforms. 
Nevertheless, a large amount of training data is needed to construct these complex systems because a lack of it could result in poor generalization performance on unseen data conditions.

Emotions can also be captured using text information.
Several techniques have been developed to represent and model this type of data.
In many Natural Language Processing (NLP) applications, word and sentence embeddings (such as Word2Vec \cite{EEO-2013} or BERT-based \cite{BPO-2018}) have proven to be the most efficient representations, yielding significant improvements in deep learning models performance.
Despite offering an enhanced representation of textual information, these techniques have certain limitations when applying them effectively to different tasks.
One of the principal difficulties is that training these models requires huge amounts of data and computational power.

The Odyssey2024 Speech Emotion Recognition Challenge \cite{OSE-2024} was organized as part of the Odyssey 2024 Speaker and Language Recognition Workshop to compare different emotion recognition systems submitted by teams all around the world.
The challenge consists of two independent tasks allowing to compare different emotion recognition systems submitted by teams all around the world.
The first task objective is to classify speech segments across the eight categorical emotional classes provided: anger, happiness, sadness, fear, surprise, contempt, disgust and a neutral state.
The second task consists of predicting emotional attributes for arousal (calm to active), valence (negative to positive), and dominance (weak to strong).
The challenge uses recordings from the MSP-Podcast corpus, which contains speech segments obtained from audio-sharing websites \cite{BNE-2017}.

In \cite{SMH-2019}, the authors proposed a sub-vector-based multi-head attention pooling to efficiently obtain discriminative speaker embeddings given non-fixed-length speech utterances.
In this system, a Convolutional Neural Network (CNN) encodes short-term speaker features from the spectrogram, which are split into homogeneous sub-vectors. 
Each self-attention pooling head is applied over a split of sub-vectors.  
This mechanism allows the model to attend to the most informative patterns from different positions of the encoded sequence.
In a later work \cite{DMH-2021}, the authors present the Double Multi-Head Attention (DMHA), where a second self-attention pooling is applied.
This additional pooling layer allows the model to select which head context vectors are the most relevant to produce an enhanced utterance-level vector.
Since this network is trained as a speaker classifier, it can be easily adapted to solve Speaker Classification tasks such as SER \cite{SCB-2022}.

In this paper we describe our Double Multi-Head Attention Multimodal System.
We used this architecture for the Odyssey 2024 Speech Emotion Recognition Challenge categorical recognition task.
Since self-supervised pre-trained models learn universal representations from massive unlabeled speech data and adapt effectively across various downstream tasks, several pre-trained versions of these architectures were explored to extract acoustic features.
Additionally, speech utterances were automatically transcribed with Whisper \cite{RSR-2023} and then text features were extracted using a pre-trained BERT \cite{BPO-2018} model.
A Multi-Head Attention layer is used to transform these speech and text features into contextualized representations.
Using this attention technique allows the model to learn complex relationships between acoustic and linguistic information.
These contextualized representations are then pooled into an utterance-level vector using a second attention mechanism.
Finally, we used a set of fully connected layers to classify the corresponding emotions.
Models were trained by applying an on-line data augmentation process.  
Three different models were used in a hard-voting ensemble to achieve the third position in the Task 1 Challenge ranking, where 31 teams participated in total.

The rest of this paper is structured as follows. 
In Section 2 we describe several studies that are related to our work.
In Section 3 Double Multi-Head Attention is explained.
Section 4 details our proposed system.
In Section 5 experimental setups and results are included. 
Concluding remarks are given in Section 6.

\section{Related Work}

Deep Neural Networks have demonstrated remarkable success in extracting emotions from speech signals in the last years \cite{MER-2022, ASE-2021, ANE-2021, FCP-2020, IET-2019}.
Commonly, a SER system includes a feature extraction component and a pooling layer that aggregates the extracted features into a single utterance-level vector.
Finally, a classification layer takes the pooled vector as input and outputs each emotion probability.
Nevertheless, the automatic recognition of emotions from speech in realistic domains remains a significant challenge, primarily due to the intricate nature of expressive behaviors that manifest during human interactions.

One major obstacle in SER is still the lack of substantial amounts of labelled data needed to construct complex enough deep learning models \cite{ERI-2018}.
To address this problem, several data augmentation methods were developed to generate synthetic data \cite{DAU-2019, OTF-2020}.
Additionally, unsupervised or self-supervised pre-training of neural networks emerged as an effective technique for settings where labelled data is scarce. 
The key idea is to learn general representations in a setup where substantial amounts of data are available.
These learned representations are then used to improve performance on a downstream task for which the amount of data is limited. 
This is particularly interesting for tasks like SER, where a substantial effort to obtain labelled data is required.

SER systems take audio signals as input.
In order to improve the performance of these systems, a feature extraction process is usually applied.
In this process, features are extracted from the audio signal by changing the speech waveform to a form of parametric representation.
Several options are available to parametrically represent the speech signal for the recognition process. 
The Mel-frequency Cepstral Coefficients (MFCC) feature extraction strategy is recognized as one of the most effective and universally adopted techniques in the speech recognition domain \cite{ASO-2021}.
Nevertheless, there are no guarantees that these hand-crafted features are optimal for all speech-related tasks and they might limit the potential of the powerful representation of DNN systems. 

To mitigate this drawback, some works aim to extract learnable acoustic features from the raw audio waveforms. 
In \cite{W2V-2020}, the wav2vec2.0 architecture quantizes continuous speech data into discrete units automatically and then guesses the correct discrete units at randomly masked locations using Transformers \cite{AIA-2017}.
XLS-R \cite{XSS-2021} is a large-scale model for cross-lingual speech representation learning based on wav2vec2.0.
Since it covers a wide range of languages, data regimes and domains, it shows a remarkable generalization capacity.
Hidden-Unit BERT (HuBERT) \cite{HSS-2021} is another self-supervised speech representation learning approach, which utilizes an offline clustering step to provide aligned target labels for a BERT-like prediction loss.
WavLM \cite{WLS-2022} jointly learns masked speech prediction and denoising in pre-training.
Some inputs are simulated noisy/overlapped speech with masks and the target is to predict the pseudo-label of the original speech on the masked region like HuBERT.
By doing this, wavLM learns universal speech representations from massive unlabeled speech data and adapts effectively across various speech processing tasks.

Regarding text feature extraction methods, word embeddings have become the standard representation in many natural language processing tasks, being Word2Vec \cite{EEO-2013} and GloVe \cite{GGV-2014} some of the most widely adopted.
Both unsupervised models have shown great success in a range of NLP tasks such as sentiment analysis, document indexing, and topic model analysis.
Nevertheless, a significant drawback of these models is that they ignore word order, which results in a loss of syntactic and semantic information of words in sentences.
By extracting representations from text data to capture the context, BERT \cite{BPO-2018} models are used to solve this limitation.
BERT stands for Bidirectional Encoder Representations from Transformers and is a language model trained on large amounts of unlabelled text data that has achieved state-of-the-art results on many NLP tasks.
In order to enhance the performance of emotion recognition, several studies have effectively combined speech-based representations with BERT-based embeddings \cite{ERB-2021, FAF-2020, MER-2022}.

Pooling strategies can be classified into two main classes: statistical and learning-based pooling.
Within the statistical pooling strategies, the most commonly used are average pooling and statistical pooling \cite{MER-2022, LDF-2018}.
These strategies assume that all elements of the pooling component input vectors contribute equally to the utterance-level vector creation.
Since every frame may not provide equal speech discriminative information, many works applied self-attention mechanisms for weighted learning-based pooling layers.
Single-head attention pooling with different self-attentive scoring functions was explored in \cite{SER-2019, IET-2019, SAN-2023}. 
Also, multi-head attention pooling approaches were applied successfully in several works \cite{MRM-2020, SAS-2018, DMH-2021}.
Because pooling layers behave differently with different datasets, network structures or loss functions, it is difficult to conclude which one is the best, if there is one. 
Pooling functions with learnable parameters achieved better results than the simple pooling layers such as the average pooling and statistical pooling in most cases, with a weakness of higher computational complexity than the latter.

Multimodal emotion recognition (MER) aims to identify human emotional states by combining different data modalities such as text, speech, images or videos.
With the advance of deep learning technologies and the increasing availability of multimodal datasets many MER systems have been explored \cite{ASO-2023, MER-2021}. 
The expressiveness of unimodal information is naturally restricted, so it frequently fails to fully convey the wide range of human emotions.
Since spoken data consists of audio and text information, studies have explored the combination of acoustic features and linguistic content for emotion recognition.
The fusion of diverse types of signals to obtain complementary information is key to improving model performance in MER.
These approaches implement this combination by mixing either each modality embeddings (known as early fusion) or decision scores (late fusion) \cite{MMF-2023, ERB-2021, FAF-2020, MER-2022}.  

\section{Double Multi-Head Attention} \label{Double Multi-Head Attention}

Attention mechanisms have become a standard component of deep learning networks, contributing to remarkable results improvements \cite{ARO-2021}.
In \cite{AIA-2017}, a Multi-Head Attention (MHA) mechanism was proposed.
Instead of performing a single attention function, first the queries, keys and values are projected $H$ times with different learned linear projections.
Then, an attention function is applied over each of these projected versions of queries, keys and values.

In \cite{SMH-2019}, the authors used a sub-vector-based MHA pooling as an efficient mechanism to obtain discriminative speaker embeddings from speech utterances.
Frame-level features are split into $H$ sets of equally-sized sub-vectors and an attention pooling is applied over each set of sub-vectors. 
The attention performed over each set of sub-vectors is a dot-product attention, where the keys and the values projections are the sub-vectors and there is only one trainable query.
In contrast with standard MHA \cite{AIA-2017}, this sub-vector MHA outputs $H$ pooled vectors and significantly reduces the number of learnable parameters.

In the architecture designed in \cite{SMH-2019}, the authors extracted speech features from the spectrogram using a CNN component.
In this paper we extract features directly from the raw waveform with pre-trained self-supervised models, allowing to remove the CNN component.
Combining this removal with a Data Augmentation process enables to significantly increase the number of parameters of the model without losing its generalization ability beyond the training data.
Motivated by this, we experiment replacing the efficient sub-vector MHA \cite{SMH-2019} layer with a standard MHA \cite{AIA-2017} layer, allowing the model to capture more complex relationships and patterns.

In a later work \cite{DMH-2021}, the authors presented the Double MHA (DMHA) approach.
After a first sub-vector-based MHA layer, a second attention layer is applied.
This additional layer uses an attention pooling mechanism over the first layer output vectors by performing a dot-product attention, where the keys and the values projections are the vectors and there is only one trainable query.
This second layer allows the model to select which vectors are the most relevant to produce the final pooled vector.

Standard and sub-vector variants for the first attention layer of the DMHA are formulated in subsections \ref{Standard Multi-Head Attention} and \ref{Sub-vector Multi-Head Attention}.
The second attention layer of the DMHA architecture is explained in subsection \ref{Attention Pooling}.

\subsection{Standard Multi-Head Attention} \label{Standard Multi-Head Attention}

Given a sequence of hidden-states $\lbrace h_t \in \mathbb{R}^D \mbox{ }| \mbox{ } t=1,...,T \rbrace$, the output of a $H$-headed standard MHA \cite{AIA-2017} layer are $T$ vectors.
These are computed using learnable matrices $W_j^Q$, $W_j^K$, $W_j^V \in \mathbb{R}^{D \times D}$ for each head $j$, and $W^O \in \mathbb{R}^{H D \times D}$:
\begin{equation}
   \mbox{MHA} (X) = \mbox{concat} \left( \mbox{head}_1, \dots, \mbox{head}_H \right) W^O,
  \label{eq1}
\end{equation}
where $X \in \mathbb{R}^{T \times D}$ is the result of packing together the input sequence $h_1, ..., h_T$ and 
\begin{equation}
   \mbox{head}_j = \mbox{softmax} \left( \frac{X W_j^Q (X W_j^K)^T}{\sqrt{D}} \right) X W_j^V.
  \label{eq2}
\end{equation}
This mechanism, which we refer to as standard MHA (sometimes also referred to as Global MHA), allows the model to jointly attend to information from different representation subspaces at different positions. 

\subsection{Sub-vector Multi-Head Attention} \label{Sub-vector Multi-Head Attention}

Given a sequence of hidden-states $\lbrace h_t \in \mathbb{R}^D \mbox{ }| \mbox{ } t=1,...,T \rbrace$, the output of a $H$-headed sub-vector MHA \cite{SMH-2019} layer are $H$ vectors.
Each hidden state $h_t$ is split into $H$ new hidden states so that $h_t = [h_{t1}, ..., h_{tH}]$, where $h_{tj} \in \mathbb{R}^{D/H}$.
Now, for each head $j$, self-attention is applied over $[h_{1j}, ..., h_{Tj}]$.
Each head weights are defined as:
\begin{equation}
  w_{tj} = \frac{\text{exp} \left( \frac{h_{tj}^T u_j}{\sqrt{d_h}} \right)}{\sum_{l=1}^{T}  \text{exp} \left( \frac{h_{lj}^T u_j}{\sqrt{d_h}} \right)},
  \label{eq3}
\end{equation}
where \( w_{tj} \) corresponds to the attention weight of the head \( j \) on the step \( t \) of the sequence, \( u_j \in \mathbb{R}^{D/H} \) is a trainable parameter and \( d_h \) corresponds to the hidden state dimension \( D/H \).
We then compute a pooled representation $c_{j}$ for each head: 
\begin{equation}
  c_j = \sum_{t=1}^{T} w_{tj}^{'} h_{tj} 
  \label{eq4}
\end{equation}

Each self-attention pooling for the head $j$ is a dot-product attention over $[h_{1j}, ..., h_{Tj}]$, where the keys and the values are both $[h_{1j}, ..., h_{Tj}]$ and there is only one trainable query $u_j$.
With this design, this mechanism allows the model to efficiently attend to the most informative patterns from different positions of the hidden states sequence, significantly reducing the number of learnable parameters.

\subsection{Attention Pooling} \label{Attention Pooling} 

Let $\lbrace c_l \in \mathbb{R}^C \mbox{ }| \mbox{ } l=1,...,L \rbrace$ be the output vectors of the first attention layer.
If standard MHA is used in the first layer, $L$ is equal to $T$; if sub-vector MHA is used, $L$ is equal to $H$.
Self-attention is now used to pool these vectors to obtain an overall utterance-level vector $c$:

\begin{equation}
  w_i^{'} = \frac{\text{exp} \left( \frac{c_{i}^T u^{'}}{\sqrt{C}} \right)}{\sum_{l=1}^{L} \text{exp} \left( \frac{c_{l}^T u^{'}}{\sqrt{C}} \right)}
  \label{eq5}
\end{equation}
\begin{equation}
  c = \sum_{l=1}^{L} w_l^{'} c_{l} 
  \label{eq6}
\end{equation}
where \( w_{i}^{'} \) corresponds to the aligned weight of $c_i$ and \( u^{'} \in \mathbb{R}^{C} \) is a trainable parameter.

With this method, each vector $c$ is computed as a weighted average of vectors, allowing the system to learn the relevance of each of these vectors for each utterance.  

\section{System Description}

\begin{figure}[t]
\includegraphics[scale=0.075]{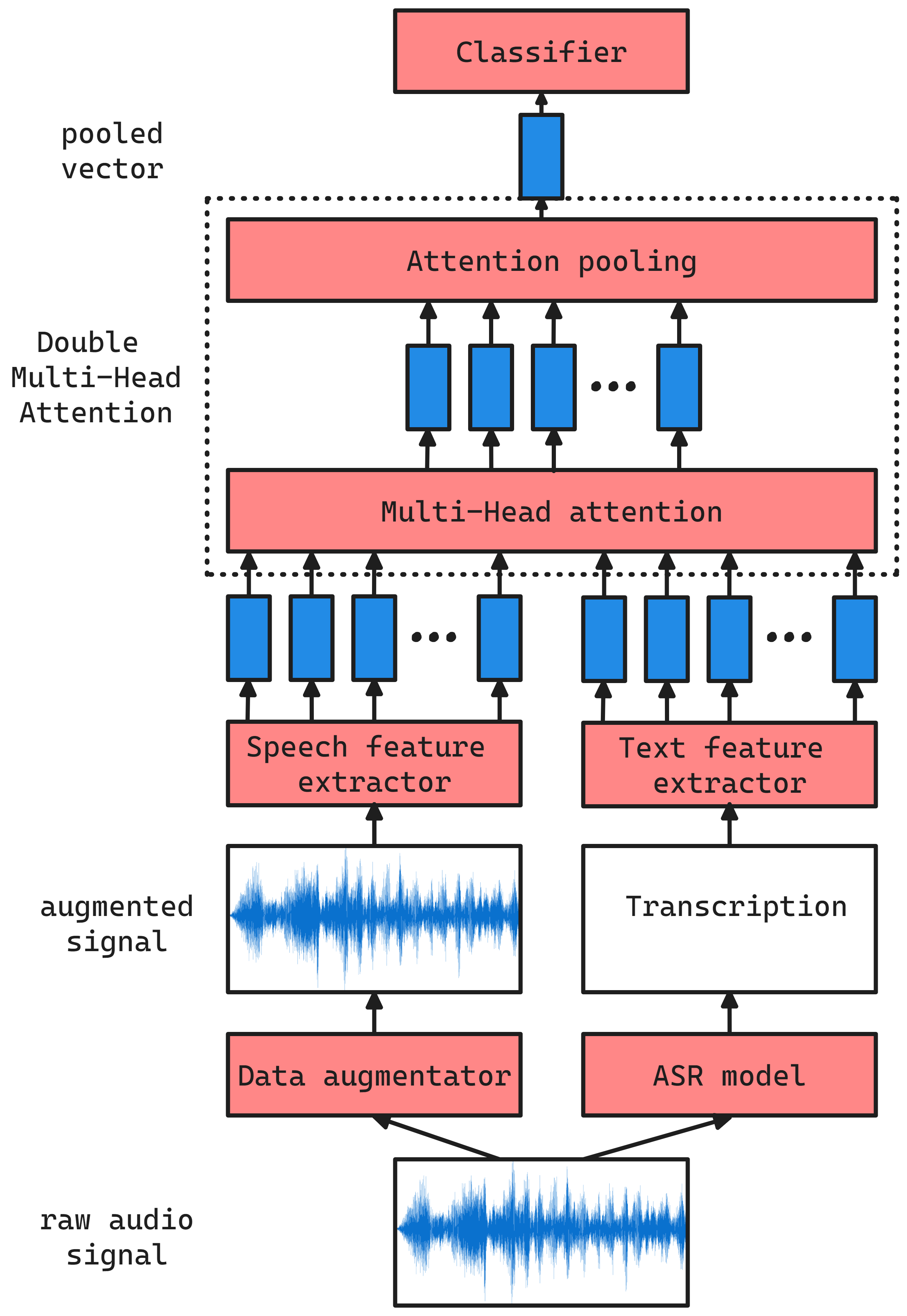}
\caption{{\it Double Multi-Head Attention Multimodal System architecture.}}
\label{fig_arch}
\end{figure}

Figure \ref{fig_arch} illustrates the architecture of our proposed multimodal emotion-recognition framework. 
The following subsections describe the components used in the Double Multi-Head Attention Multimodal System architecture.

\subsection{Speech Features}

Acoustic features were extracted directly from the waveform using pre-trained self-supervised models. 
Some of the most widely used and best-performing models to that end are wav2vec2.0 \cite{W2V-2020}, XLS-R \cite{XSS-2021}, HuBERT \cite{HSS-2021} and wavLM \cite{WLS-2022}. 
These pre-trained models get speech features representations using several Transformer layers. 
Every model has different versions with variations in the number of Transformer layers, the dimension of the speech representations and the training data.
We experimented with the following:
\begin{itemize}
    
    \item wav2vec2.0: wav2vec2.0 model ("large-lv60k" architecture), pre-trained on 60,000 hours of unlabeled audio, not fine-tuned.
    
    \item XLS-R: XLS-R model with 300 million parameters, pre-trained on 436,000 hours of unlabeled audio from multiple datasets in 128 languages, not fine-tuned.
    
    \item HuBERT: HuBERT model ("large" architecture), pre-trained on 60,000 hours of unlabeled audio, not fine-tuned.
    
    \item wavLM: wavLM Large model ("large" architecture), pre-trained on more than 80,000 hours, not fine-tuned.
\end{itemize}

Every one of these versions uses 24 Transformer layers to get 1024-dimensional speech representations. 
Since every Transformer layer may produce a different level of complexity, a multi-level aggregation is often suggested \cite{SSP-2021}.
To explore this approach, we extract the acoustic features of each model performing a weighted sum of every Transformer layer, where the weights were learned by the model at the training stage.

\subsection{Text Features}

To extract text features from the waveforms we first used Whisper \cite{RSR-2023} to get audio transcriptions.
Whisper is an Automatic Speech Recognition Transformer-based system trained with up to 680,000 hours of multilingual and multitask data.
When compared to humans, this model approaches their accuracy and robustness.
A pre-trained BERT model was used to extract text features from the transcriptions.
The BERT large uncased pre-trained version was used, which was trained on lower-cased English text. 
This version has a total of 340M parameters, using 24 Transformer layers to output 1024-dimensional text representations.
In this case, only the last Transformer layer output was used.

\subsection{Double Multi-Head Attention Multimodal Fusion}

Given a speech segment, the input of the DMHA Multimodal Fusion component is the acoustic and text features.
Several studies have explored the combination of acoustic features and linguistic content for emotion recognition, concluding that the fusion of diverse types of signals is key to improving model performance \cite{ASO-2023}.
Motivated by this, these acoustic and text features were mixed into a first MHA layer to let the model learn complementary information using self-attention scores.
Then, a second attention layer is applied to generate an utterance-level pooled vector.

If there are $T_1$ acoustic features and $T_2$ text features, the $T = T_1 + T_2$ features are input to the first MHA layer.
This first attention layer can consist of a sub-vector MHA or a standard MHA as described in section \ref{Double Multi-Head Attention}.
If a $H$-headed sub-vector MHA is used, this first layer will output $H$ vectors.
In the case of a $H$-headed standard MHA, this first layer will output $T$ vectors.
We experimented with 4-headed sub-vector and standard MHA layers.
This first MHA layer outputs multimodal contextualized representations which are then aggregated into a single utterance-level vector using a second attention layer with an attention pooling strategy as described in section \ref{Attention Pooling}. 

\subsection{Classification Layer}

The aggregated utterance-level vector $c$ obtained from the pooling component is fed into a classification layer.
This component consists of a set of fully connected layers: an input layer, several hidden layers and an output layer. 
The input layer has the same width as $c$'s dimension. 
The output layer has the same width as the number of classes of the task and a SoftMax layer is applied to get each class probability.

\section{Experiments}

\subsection{Database}

We evaluate the effectiveness of the proposed multimodal emotion recognition system on the MSP-Podcast dataset \cite{BNE-2017}, which
contains speech segments obtained from audio-sharing websites. 
The speaking turns have been perceptually annotated by at least five raters with categorical and attribute-based emotional labels.
Categorical eight emotional classes provided in the dataset are happiness, sadness, fear, surprise, contempt, disgust and a neutral state.
While the distribution for attribute-based emotional labels has balanced content, the distribution for emotional categories is less balanced.
The training set has 68,119 speaking turns. 
The development set has 19,815 speaking segments from 454 speakers. 
The test set comprises 2,347 unique segments from 187 speakers, for which the labels have not been made publicly available for the challenge.

It has been proven that using Data Augmentation techniques to increase the volume of the training data significantly improves the performance of speech recognition systems \cite{OTF-2020}.
An online Data Augmentation process \cite{OTF-2020} was used applying speaker augmentation with speed perturbation \cite{SAN-2019}, room impulse responses (RIRs) from \cite{ASO-2017} and background noises from the MUSAN database \cite{MAM-2015} directly to the waveform.

\subsection{Experimental Setup}

Raw waveforms were normalized using their overall mean and standard deviation from the training dataset.
We trained our models using batches of size 32. 
Every speech signal was randomly cropped with a 5.5-second window, given that the median of the training utterances length is 5.2 seconds.
When needed, a repetition padding was used.
Each waveform was augmented with a $0.5$ probability.
In that case, one augmentation technique was randomly chosen from speed perturbation, RIRs or background noises.
Full-length waveforms without augmentation were used at the evaluation stage.

Acoustic and text features were extracted using frozen self-supervised pre-trained models.
Each feature has a dimension equal to 1024.
Pre-computed Whisper transcriptions were used before extracting text features.
A 4-headed Double Multi-Head Multimodal Fusion was used as the pooling component.
We experimented using both standard and sub-vector-based MHA layers.
Finally, the 8-emotion classification was done by passing the pooled vector through the classification component.
The input layer of the classification component has the same width as the aggregated utterance-level vector dimension obtained from the pooling component. 
The hidden layers of the classification component are a set of 4 512-dimensional dense layers.
Each of the input and hidden layers is followed by a Layer Normalization \cite{LN-2016}, Gaussian Error Linear Units (GELU) activations \cite{GEL-2016} and a $0.1$ probability of dropout.

Since the F1-score (macro) was the metric defined to evaluate the categorical emotions prediction task (task 1 from the Challenge), training and validation phases were monitored and evaluated with this score.
Each team had a maximum of three submissions per task and the submitted systems were evaluated in the test set, from which labels were not publicly available.
Every model was trained for 20 epochs, using early stopping to avoid overfitting.
All our models were trained and evaluated using 2 GPUs (NVIDIA GeForce GTX TITAN X, NVIDIA TITAN Xp or NVIDIA GeForce RTX 2080 Ti).
Every 20 epochs training took around 24 hours to finish.
AdamW \cite{DWD-2017} was selected as the optimizer.
Different learning rates were tried, using a 50\% decay every 5 epochs without validation of F1-score improvement.

Regarding loss functions, we experimented with Weighted Cross-Entropy (WCE) Loss and Focal Loss \cite{FLF-2017} since they both act as effective alternatives to deal with class imbalance.
The training inverse frequency of each class was used as weights for the WCE Loss.
The Focal Loss function is a dynamically scaled cross-entropy loss, where the scaling factor decays to zero as confidence in the correct class increases. 
Intuitively, this scaling factor can automatically down-weight the contribution of easy examples during training and rapidly focus the model on hard examples.
This scaling factor is governed by a parameter gamma, which was set to $2$.

Once each model was trained, a per-class threshold adjustment was performed.
The idea is to adjust each class decision threshold to improve their precision and recall.
For each class, we convert the task to a binary classification by treating the rest as a whole one negative class.
Then, different thresholds were explored to maximize the training Macro F1-score.
Once every threshold is set, the way of predicting the final class is the following: if the class with the highest probability output surpass its corresponding threshold, then that class is predicted (no change is done).
If the class with the highest probability output do not surpass its corresponding threshold, then the class with the next highest probability is predicted.

After the training phase and threshold adjustments were applied, we experimented with mixing different models to get a 3-model ensemble.
The motivation for using ensemble models is to reduce the generalization error of the final prediction. 
As long as the base models are diverse enough, the prediction error of the ensembled model can decrease.
A hard voting strategy was adopted for the ensemble.
In case of a tie, we used the prediction of a predefined model (the one with the best validation Macro F1-score).

\begin{table}[bh]
\vspace{2mm}
\centerline{
    \renewcommand{\arraystretch}{1.5}
    \begin{tabular}{|c|c|p{1.5cm}|p{1.5cm}|}
        \hline
        \multicolumn{2}{|c|}{\multirow{2}{*}{\textbf{Configuration}}}        & \textbf{Train Macro F1-score}  & \textbf{Validation Macro F1-score} $\downarrow$ \\
        \hline
        \multicolumn{2}{|l|}{Ensemble of models}          & $34.88\%$      & $33.80\%$ \\
        \hline
        WCE Loss            & XLS-R                                  & $34.42\%$      & $33.43\%$ \\
        \hline
        Focal Loss          & XLS-R                                  & $39.15\%$      & $33.37\%$ \\     
        \hline
        WCE Loss & wav2vec2.0   &  $34.83\%$  & $32.69\%$ \\ 
        \hline
        Focal Loss & HuBERT        &  $37.84\%$  & $32.40\%$ \\
        \hline
        WCE Loss & HuBERT            &  $36.73\%$  & $32.18\%$ \\
        \hline
        WCE Loss & wavLM          &  $32.48\%$  & $31.44\%$ \\
        \hline
        Focal Loss & wav2vec2.0   &  $35.75\%$  & $31.27\%$ \\
        \hline
        Focal Loss & wavLM            &  $33.24\%$  & $30.77\%$ \\
        \hline
        \multicolumn{2}{|l|}{Challenge Official Baseline}               &  - & $30.70\%$ \\
        \hline
    \end{tabular}
}
\caption{\label{table_standard_mha} {\it Experimental results using the standard MHA mechanism in the DMHA component. 
Threshold adjustment was applied to every model, except for the Challenge Official Baseline.
Ensemble of models combine the following three models: WCE Loss and XLS-R; WCE Loss and wav2vec2.0; WCE Loss and wavLM.}}
\end{table} 

\subsection{Results}

We trained several models combining different loss functions (WCE loss and Focal loss), pre-trained self-supervised speech features extractors (wav2vec2.0, XLS-R, HuBERT and wavLM) and MHA mechanisms (standard and sub-vector MHA).
A model ensemble using the following three configurations was also evaluated: WCE loss and XLS-R speech features extractor; WCE loss and wav2vec2.0 speech features extractor; WCE loss and wavLM speech features extractor.
The choice of these three models was motivated by the idea that extracting different patterns from the same speech utterance may contribute to the final system diversity.
After applying threshold adjustment, we evaluate them using Macro F1-score both in the training and validation datasets of the competition.
Applying threshold adjustment increased validation Macro F1-score approximately in a 1\% absolute both for the Standard and the sub-vector-based models, being a useful strategy for this competition.

We can see from Table \ref{table_standard_mha} results that our best model is the ensemble of models. 
This configuration was used to obtain our best submission ranking at the Challenge.
This model outperforms the challenge official baseline with a 3.1\% absolute increase in the validation Macro F1-score.
In terms of relative improvement, our best model gained 10.1\% validation Macro F1-score compared to the baseline.
When using standard MHA, models with XLS-R \cite{XSS-2021} as speech feature extractor were the ones with the best performance.

\begin{table}[h]
\vspace{2mm}
\centerline{
    \renewcommand{\arraystretch}{1.5}
    \begin{tabular}{|c|c|p{1.5cm}|p{1.5cm}|}
        \hline
        \multicolumn{2}{|c|}{\multirow{2}{*}{\textbf{Configuration}}} & \textbf{Train Macro F1-score} & \textbf{Validation Macro F1-score} $\downarrow$ \\
        \hline
        WCE Loss & HuBERT    &  $37.88\%$  & $30.84\%$ \\
        \hline
        Focal Loss  & XLS-R      &  $37.41\%$  & $30.73\%$ \\
        \hline
        \multicolumn{2}{|l|}{Challenge Official Baseline}               &  - & $30.70\%$ \\
        \hline
        WCE Loss & XLS-R      &  $36.05\%$  & $30.53\%$ \\
        \hline
        Focal Loss  & wav2vec2.0  &  $31.20\%$  & $29.31\%$ \\
        \hline
        WCE Loss & wav2vec2.0  &  $30.29\%$  & $28.93\%$ \\
        \hline
        Focal Loss  & HuBERT    &  $31.94\%$  & $28.89\%$ \\
        \hline
        Focal Loss  & wavLM     &  $32.28\%$  & $27.67\%$ \\
        \hline
        WCE Loss & wavLM     &  $31.69\%$  & $26.13\%$ \\
        \hline
    \end{tabular}
}
\caption{\label{table_subvector_mha} {\it Experimental results using the sub-vector MHA mechanism in the DMHA component. 
Threshold adjustment was applied to every model, except for the Challenge Official Baseline.}}
\end{table}

\begin{figure}[h]
\includegraphics[width=\columnwidth]{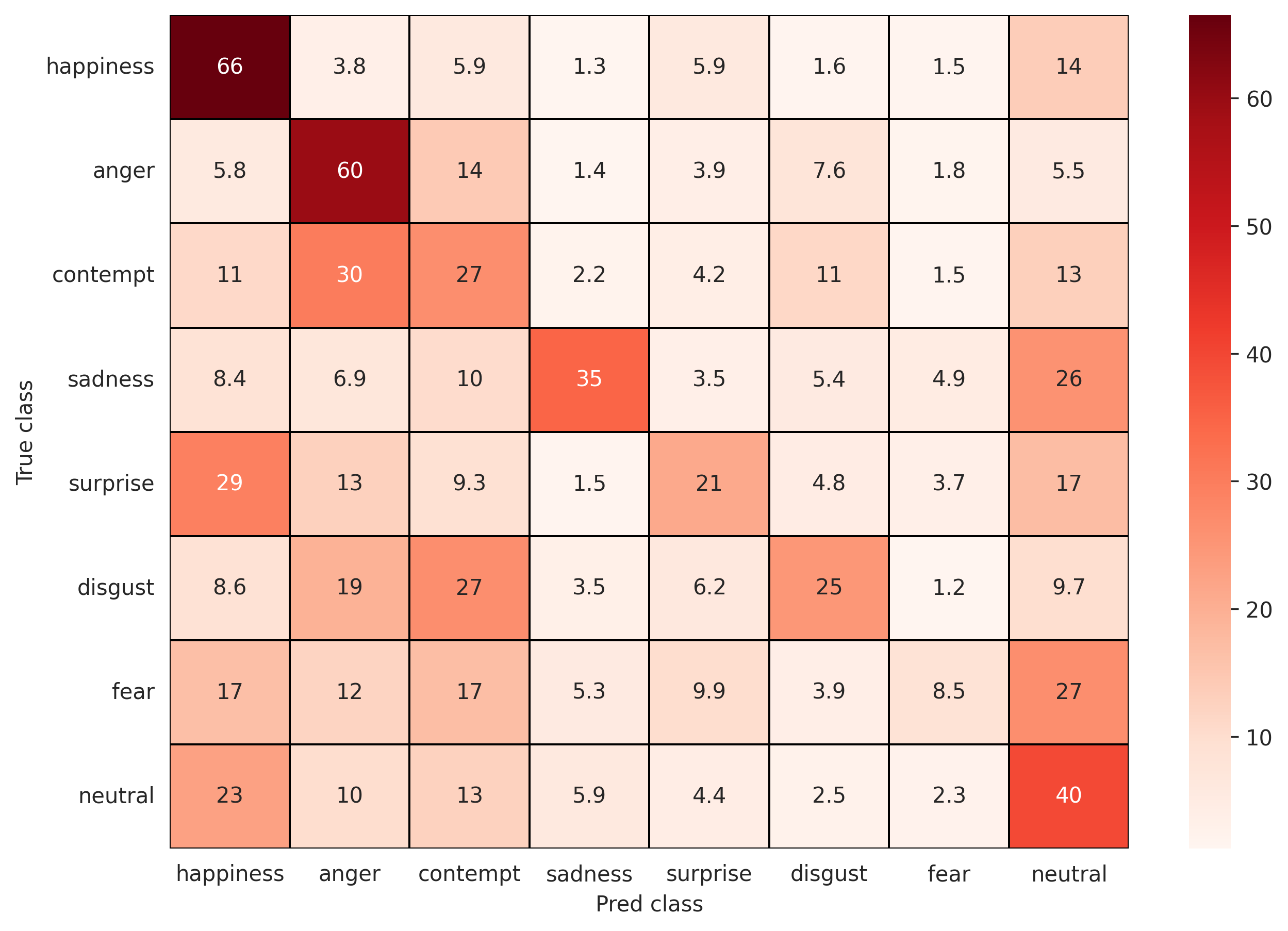}
\caption{{\it Validation confusion matrix of the ensemble model.}}
\label{confusion_matrix}
\end{figure}

Figure \ref{confusion_matrix} shows the validation confusion matrix for the ensemble model.
We can see that the system achieves the best results detecting happiness and anger.
There are some emotions that may be more similar and the model can confuse: when the real emotion is contempt, it predicts anger 30\% of the time; a similar situation arises for surprise and happiness; disgust is predicted as anger or contempt around 45\% of the cases.
Finally, fear is the worst performance class.

\begin{figure}[H]
\includegraphics[width=\columnwidth]{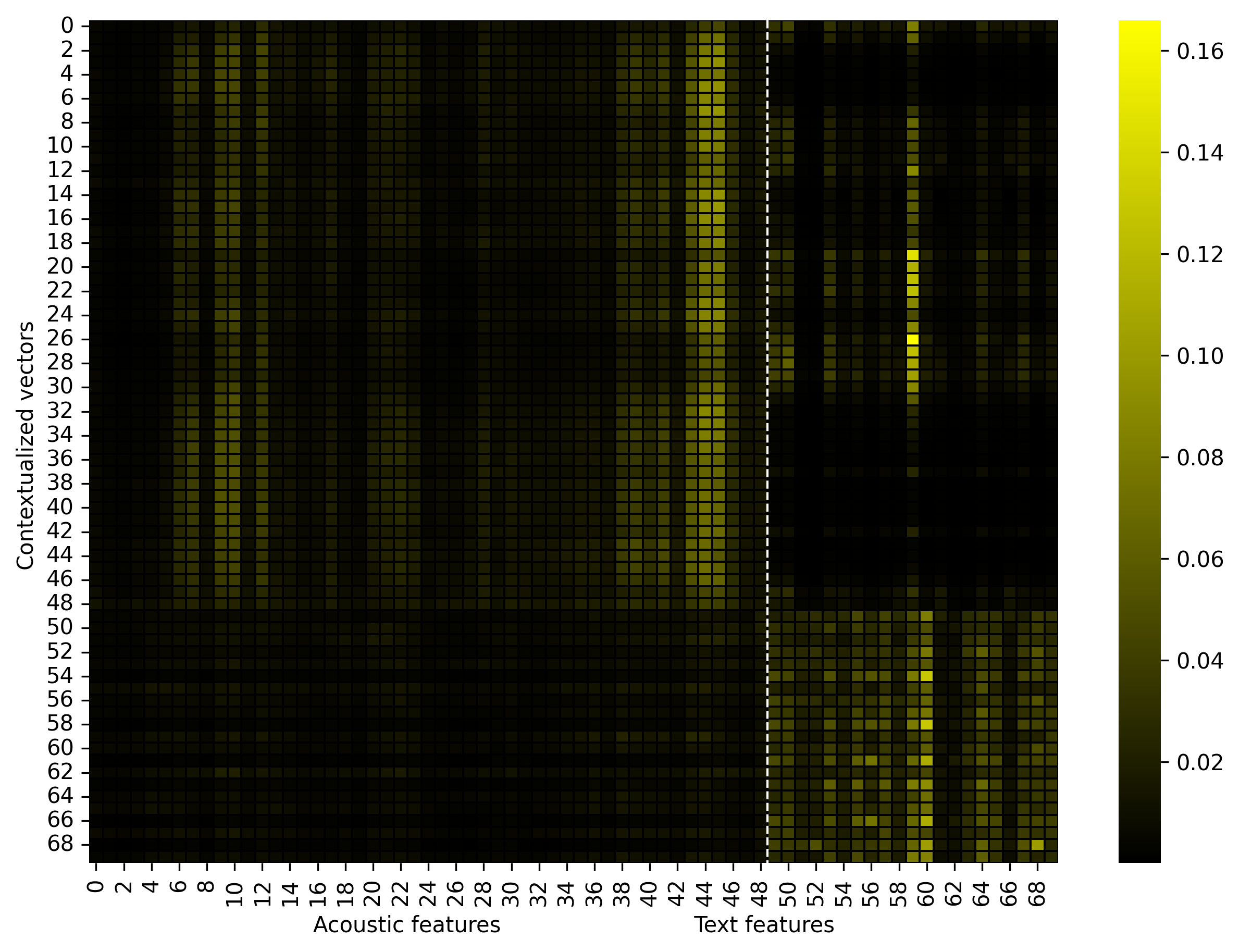}
\caption{{\it Head 0 attention weights visualization. The white vertical dashed line divides speech features from text features.}}
\label{att_mixed}
\end{figure}

\begin{figure}[H]
\includegraphics[width=\columnwidth]{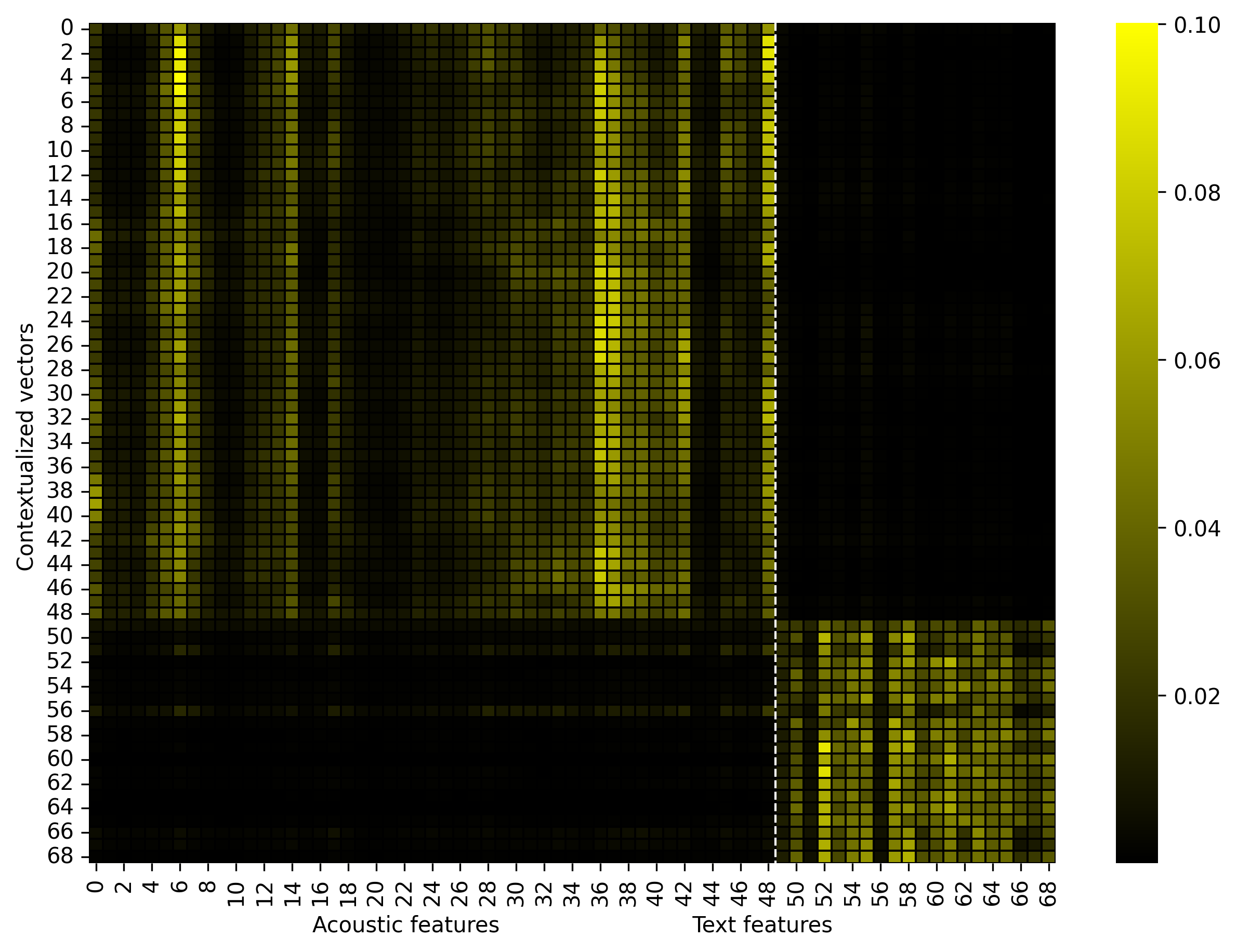}
\caption{{\it Head 1 attention weights visualization. The white vertical dashed line divides speech features from text features.}}
\label{att_separated}
\end{figure}

We used the trained model corresponding to the WCE loss and XLS-R speech feature extractor configuration with standard MHA to analyze the features fusion in the MHA layer.
To that end, each head attention weights of a particular audio sample were calculated.
For each head, every contextual representation uses learned attention weights (matrix rows) to attend to each of the speech and text features (matrix columns), as visualized in figures \ref{att_mixed} and \ref{att_separated}.
Figure \ref{att_mixed} shows that, for head 0, attention is focused both on some speech and text features (specially in rows 20 to 30) to generate contextual representations.
On the other hand, in figure \ref{att_separated} we can see that, for head 1, some contextual representations attend mostly to speech features and others mostly to text features.
This shows how our Multi-Head Attention Multimodal Fusion creates complex contextual representations combining speech and text information in different ways.

The results of the experiments using the sub-vector MHA approach are shown in Table \ref{table_subvector_mha}.
Given that both the configurations with WCE Loss and HuBERT speech feature extractor and with Focal Loss and XLS-R speech feature extractor surpass the baseline validation score, we conclude that combining a powerful self-supervised feature extractor with a sub-vector-based Double Multi-Head Attention Multimodal Fusion component can obtain good results for this task.
Since the sub-vector MHA has significantly less parameters than the standard MHA, this efficient architecture could be useful in settings where large models are not allowed and/or training data is scarce, which is a common scenario in SER.

\section{Conclusion}

In this paper we have described our Double Multi-Head Attention Multimodal System for the Odyssey 2024 Speech Emotion Recognition Challenge.
Acoustic and text features were extracted using pre-trained self-supervised models.
These multimodal features are mixed adopting an early fusion strategy.
First, an MHA layer generates complementary contextualized representations.
A second attention layer is then applied to pool these representations into an utterance-level vector.
For the first attention layer, we experimented with different mechanisms: standard MHA and sub-vector MHA.
Since the sub-vector variant has significantly less parameters than the standard one, our obtained results show that this efficient architecture could be useful in settings where large models are not allowed and/or training data is scarce.
On the other hand, applying standard MHA allows the model to capture complex relationships jointly attending to information from different representation subspaces. 
In terms of relative improvement, our best model gained 10.1\% validation Macro F1-score compared to the baseline.
This model outperforms the challenge official baseline with a 3.1\% absolute increase in the validation Macro F1-score, achieving the third position in the categorical task ranking, where 31 teams participated in total.

\section{Acknowledgments}

This work has been promoted and financed by the Generalitat de Catalunya through the Aina project and by the Spanish Ministerio de Ciencia e Innovación through the project AdaVoice PID2019-107579RB-I00.
The first author is supported by a FI grant from the Catalan government.

\bibliographystyle{IEEEbib}
\bibliography{Odyssey2024_BibEntries}

%

\end{document}